\documentclass[aps,twocolumn,showpacs]{revtex4}
\usepackage{graphicx}
\begin{document}

\title{Multifractal burst in the spatio-temporal dynamics of jerky flow}
\author{M.S. Bharathi$^{1}$, M. Lebyodkin$^{2}$, G. Ananthakrishna$^{1}$\footnote{Author for correspondence. email: garani@mrc.iisc.ernet.in},
 C. Fressengeas$^{3}$ and L.P. Kubin$^{4}$}
\affiliation{$^1$ Materials Research Centre,
Indian Institute of Science, Bangalore 560 012, India,\\
$^2$ Institute of Solid State Physics, Russian Academy of Science, 142432
Chernogolovka, Moscow district, Russia,\\
$^3$Laboratoire de Physique et M\'{e}canique des Mat\'{e}riaux,
Universit\'{e} de Metz, Ile du Saulcy, 57045 Metz Cedex 01, France,\\
$^4$Laboratoire d'Etude des Microstructures, CNRS-ONERA,
29 Av.de la Division Leclerc, 92322 Chatillon Cedex, France}

\begin{abstract}
The collective behavior of dislocations in
jerky flow is studied in Al-Mg polycrystalline samples  subjected to constant strain rate tests.
Complementary dynamical, statistical and multifractal analyses are carried
out on the stress-time series recorded during jerky flow to
 characterize the distinct spatio-temporal dynamical regimes.
It is shown that the hopping type B and the propagating type
A bands correspond to 
chaotic and self-organized critical states respectively. The crossover
between these types of bands is identified by a large spread in
the multifractal spectrum. These results are interpreted on the
basis of competing scales and mechanisms.
\end{abstract}
\maketitle


Jerky flow or the repeated yielding of alloys during plastic
deformation has been studied since the turn of the last century \cite{PLC},
 and still continues to engage the attention of
scientists. The Portevin - Le Chatelier effect (PLC), as it is
referred to, has been observed in many dilute alloys \cite{Fri64}.
It is one of the few striking examples of the complexity of the
spatio-temporal dynamics arising from {\it the collective behavior
of defect populations}. In a tension
test with a constant imposed strain rate $\dot\epsilon_a$, in
practice a constant pulling velocity $V$, the effect manifests
itself as a series of serrations in the stress-time or strain
curve. Each stress drop is associated with the nucleation  of a
band of localized plastic deformation which  under certain
conditions propagates along the sample. The continued interest in
the PLC effect arises not only from its intriguing spatio-temporal
behavior but also from its detrimental influence on  the
mechanical properties of materials. Our purpose is to
quantify the connection between the spatial (types of deformation
bands) and temporal (nature of serrations) manifestations of the
phenomenon.

It is now accepted \cite{Fri64,van75} that the microscopic
origin of the PLC effect is the dynamic aging
of the material due to the interaction between
mobile dislocations and diffusing solute atoms. In a certain range
of strain rates and temperatures, the diffusion time of the solute
atoms is of the order of the waiting time of
the dislocations temporarily arrested at  obstacles.
Then, solute atoms can diffuse to, and age the dislocations.
Thus, the stress needed
to unpin the dislocations increases with increasing waiting time and
decreasing strain rate.
At the macroscopic scale, this inverse force versus flux relation translates
into a negative sensitivity of the flow stress to the strain rate.
When the imposed strain rate falls into this range, the plastic deformation
becomes nonuniform.

In polycrystals, the metallurgical taxonomy 
distinguishes three generic types of serrations, namely types A, B
and C \cite{Micha00}. On
increasing strain rates or decreasing the temperature, one
first finds the type C, identified with randomly nucleated static
bands and large characteristic stress drops. Then
the type B with smaller serration amplitudes is found. The bands
formed  are still localized and static in nature, but forming ahead of the previous band in a spatially correlated way
giving the visual
impression of a hopping propagation. 
 Finally, one observes
continuously propagating
 type A bands associated with small stress drops. The decrease in the amplitude of the
stress drop with
increasing $\dot\epsilon_a$ reflects the decrease in the solute concentration on
the arrested dislocations.

This wealth of spatio-temporal features has long defied a proper
understanding. 
Renewed interest has come from  the connections made to nonlinear
dynamics. Dynamical \cite{Anan95,Noro97,Anan99} and statistical methods \cite{Micha00,Anan99,Leby95,Danna}
have recently been applied to characterize the complexity of
 the stress - time series. 
Studies in Refs. \cite{Micha00,Anan95,Noro97,Anan99,Leby95} suggest that two
distinct dynamical regimes encountered in single crystals might also be found in polycrystals. 
One is the chaotic dynamics \cite{Anan99}
characterized by only a few degrees of freedom,
identified as stress and dislocation densities \cite{Anan81}.
The other is self-organized critical (SOC) dynamics \cite{Anan99},
involving infinite degrees of freedom \cite{Bak87}. In polycrystals, only 
chaotic dynamics has been identified so far \cite{Noro97}.

This paper is focused on type B and A bands in polycrystalline samples as they are expected to show interesting dynamical behavior in contrast to the uncorrelated type C bands not considered here. We propose to identify the different dynamical
regimes and their correlations  with type B and A bands in
polycrystalline samples. 
 To unravel this missing
connection, dynamical, statistical and multifractal
analyses of the stress-time series are combined with the identification of the
spatial band patterns. The focus is on the multifractal
analysis. The motivation for using this method stems from a
conceptual similarity between the present transition from hopping
to propagating bands and the Anderson transition in disordered
systems \cite{Ande58}. In the Anderson model \cite{Sche91}, wave
functions  are localized when the energy, $E$, is 
below the mobility edge $E_c$
and extended for $E>E_c$. At $E=E_c$, the states are shown to
exhibit a multifractal character \cite{Sche91}. In the PLC effect,
the hopping type B bands are essentially {\it localized}, whereas
type A bands are {\it delocalized} in the sense that they are
propagating. It is this crossover from localized to delocalized
nature of the bands that we wish to capture by multifractal
analysis. We identify the dynamical regimes associated with
types B and A bands as chaotic and SOC type respectively, and 
show that the range of multifractality exhibits a sharp peak in
the transition region.  
This is shown to be a consequence of a few relevant competing scales and mechanisms.

The tensile samples of an $Al-2.5\%Mg$ alloy, of length $L=70 \,
mm$, were cut out of a polycrystalline cold-rolled sheet, parallel
to the rolling direction. The grains were anisotropic in shape,
with an aspect ratio of 5.
Tests were carried out at room temperature for eight values of
$\dot \epsilon_a=V/L$ in the range of type B and A bands, from
$5.56 \times 10^{-6} s^{-1}$ to $1.4 \times 10^{-2} s^{-1}$. The
stress was recorded at a sampling rate of 20 Hz - 200 Hz.
Three
typical stress-time, $\sigma(t)$, curves (corrected for the drift due to strain
hardening ) are shown in Fig. 1  together with $\vert d\sigma/dt \vert$.
\vspace{-0.4cm}
\begin{figure}[ht]
\includegraphics[height=7.2cm,width=8cm]{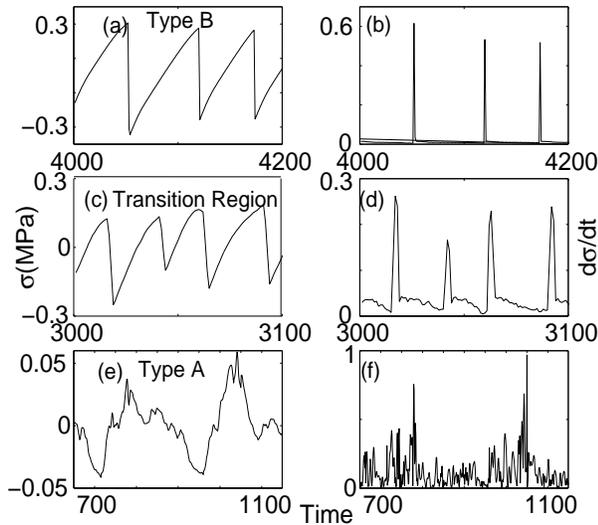}
\vspace{-0.2cm}
\caption{Stress-time series at three strain rates: (a)$5.56{\times}10^{-6} s^{-1}$, (c)$2.8 \times 10^{-4}s^{-1}$ and (e) $5.56 \times 10^{-3}s^{-1}$. (b),(d) and (f) are the corresponding plots of $\vert d\sigma/dt\vert$.}
\end{figure}

The dynamical analysis starts by embedding the time series in a
higher dimensional space using the time delay technique. Consider
the stress signal $\{\sigma(k), k=1,2,...,K\}$, where $K$ is the
number of data points and $k$ is in units of the sampling time.
Then, a $d$-dimensional vector is defined by $\vec{\xi}_k =
[\sigma(k),\sigma(k+\tau),...,\sigma(k+(d-1)\tau)], k=1,..., (K
- (d-1) \tau)$, where $\tau$ is a delay time (in units of the
sampling time). The set $\{\vec{\xi}_k, k=1,...,(K-(d-1)\tau)\}$ constitutes the
reconstructed attractor. The correlation dimension, $\nu$,  is
obtained by using the Grassberger-Procaccia algorithm \cite{Gras83}.
In this  method one calculates the correlation
integral $C(r)$,  defined as the fraction of the pairs of vectors
($\vec \xi_i$, $\vec \xi_j$) closer than a specified  value $r$.
The self similar structure of the attractor, when it exists, is
revealed by the scaling relation $C(r) \sim r^{\nu}$  in the
limit of small $r$. As the embedding dimension $d$ is increased,
the slope $ln \, C(r)/ ln \, r$ tends to a constant value
taken as the correlation dimension. The Lyapunov spectrum is
computed using Eckmann's algorithm \cite{Eckm86}, suitably modified
for short noisy time series. In our algorithm, the number of
neighbors of a vector $\vec \xi_i$ contained in a shell centered
on $\vec \xi_i$ (whose size $\epsilon_s$ is measured in $\%$ of
the size of the attractor) is large enough to properly sample the
statistics of uncorrelated noise corrupting the original signal.
 We consider the dynamics to be deterministic if a fair range of $\epsilon_s$ can be found such that
stable values (i.e. constant in that range) emerge for a positive
exponent and a zero exponent \cite{Anan99}.

In the region of strain rates $5.6 \times 10^{-6} s^{-1} \le \dot
\epsilon_a \le 1.4 \times 10^{-4} s^{-1}$, the data sets typically
contain $10000$ to $12000$ points. The band patterns are of type
B. We illustrate the results with the data file at $\dot
\epsilon_a = 5.6 \times 10^{-6} s^{-1}$. Fig. 2 shows the log-log
plot of $C(r)$ for $d = 15 $ to 18 using  $\tau =8$. The slopes
are seen to converge to a value $\nu \sim 4.6$ for $d =17$ and 18
in the range $ - 4. 9 < ln \, r < - 3.2$. 
\vspace{-0.3cm}
\begin{figure}[ht]
\includegraphics[height=3.7cm,width=6cm]{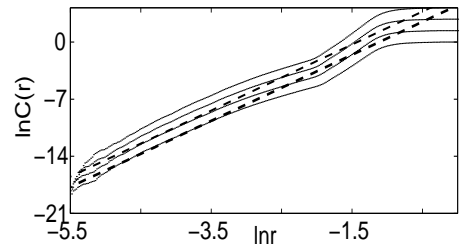}
\vspace{-0.4cm}
\caption{Correlation integral C(r) using $d=15$ to $18$, $\tau = 8$ for $\dot{\epsilon}_{a}=5.56 \times 10^{-6}s^{-1}$. The curves corresponding to $d =15$ to 17 have been displaced with respect to $d =$ 18 by a constant amount. Dashed lines are guide to the eye.}
\end{figure}
\noindent
The Lyapunov spectrum
shown in Fig. 3 has been calculated using $d = 5$. Stable positive
and zero Lyapunov exponents are seen in the range $6\% <
\epsilon_s < 12\%$. The Lyapunov dimension $D_{KY}$ obtained from
the spectrum by using the Kaplan-Yorke conjecture turns out to be
$D_{KY} \approx 4.6 = \nu$. Therefore, we conclude that
the time series is of chaotic origin. Similar results were
obtained with other samples in this region of $\dot \epsilon_a$,
with values of $\nu$ and $D_{KY}$ in the range $4.4$ to $4.6$.  In
contrast, the correlation dimension did not converge and no
positive Lyapunov exponent could be found for the data sets at
higher $\dot \epsilon_a$, implying that the dynamics is no more
chaotic in that region of strain rate. In these cases, only 
the largest Lyapunov exponent could be calculated since the data
contain only 3000 to 5000 points.
\begin{figure}[ht]
\vspace{-5.9cm}
\includegraphics{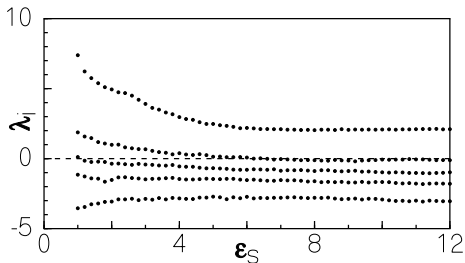}
\vspace{-1.9cm}
\caption{Lyapunov exponents vs. shell size $\epsilon_s$ for applied strain rate $\dot{\epsilon}_{a}= 5.56 \times 10^{-6}s^{-1}$; embedding dimension $d=5$.}
\end{figure}

The quantity $| d\sigma / dt| $ clearly reflects the bursts of
plastic activity (see Fig. 1). Therefore, we use its finite difference
approximant  $\psi_i(t_i)$ for statistical and
multifractal analysis. Let $\Delta \psi$ denote the amplitude of 
the bursts 
and $\Delta t$ their durations.  
We  investigate the distributions $D(\Delta \psi)$ of $\Delta\psi$,
and $D(\Delta t)$ of $\Delta t$. Plots of $D(\Delta \psi)$
are shown in Fig. 4 for appropriately chosen values of $\dot
\epsilon_a$. Peaked distributions are seen in the chaotic regime
($5.6 \times 10^{-6} s^{-1}$ to  $1.4 \times 10^{-4} s^{-1}$)
indicating the existence of characteristic values. Now we turn to higher strain rates. On the basis of single crystal results, we expect to find SOC state. In  SOC type dynamics, power law statistics arises when spatially extended driven systems  naturally evolve to a critical state. 
On increasing $\dot\epsilon_a$, the distributions become broader and
asymmetrical in the mid-region of $\Delta \psi$ eventually leading to power law distributions (Fig. 4c).
\vspace{-0.2cm}
\begin{figure}[ht]
\includegraphics[height=4cm,width=8cm]{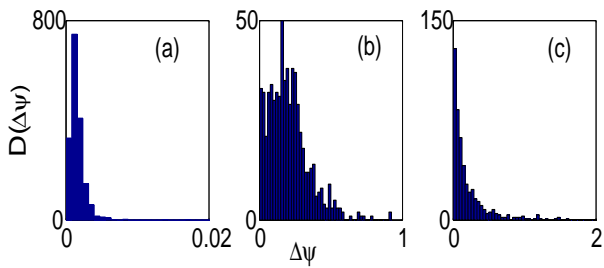}
\vspace{-0.8cm}
\caption{Distributions $|D(\Delta \psi)|$  for the data sets at (a) $\dot{\epsilon}_{a}=5.56 \times 10^{-6}s^{-1}$, (b) $1.4 \times 10^{-3}s^{-1}$ and (c) $5.56 \times 10^{-3}s^{-1}$}.
\end{figure}
\noindent
 For the data set at $\dot
\epsilon_a = 5.6 \times 10^{-3} s^{-1}$, the distribution has the form $D(\Delta\psi) \sim \Delta\psi^{-a}$ over one order of magnitude in $\Delta\psi$ with $a \sim 1.5$.
Similarly, 
we find a power law distribution $D(\Delta t) \sim \Delta t^ {-b}$
 for the duration of the bursts with $b \sim 3.2$ and for 
the conditional average $<\Delta \psi>_c  \sim \Delta t^{x}$
with $x \sim 4.2$.
 Thus, the scaling relation
$x(a -1) + 1 = b $ that characterizes SOC dynamics is well
satisfied. Similar results  are also found at the
highest strain rate $\dot\epsilon_a =  1.4 \times 10^{-2}s^{-1}$.
Thus, the region of SOC-type dynamics extends from $\dot
\epsilon_a =5.6 \times 10^{-3} s^{-1}$ onward, coinciding with the
region of type A bands.

Recall that a multifractal analysis was used to quantify the
broad 
distribution of length scales occurring in the transition region 
between localized and delocalized states in the Anderson transition \cite{Sche91}.
In our case also, there is a broad distribution of time scales 
in the region separating chaos and SOC (Fig. 4b).
Thus, we  anticipate that multifractal analysis will quantify this 
 heterogeneity \cite{Kad86}.     
Let $N = N(\delta t)$ 
be the number of time intervals  $\delta t$ with $m$ points, required to cover 
$[0,K]$. Then, the normalized amplitude of the bursts in the $i$th interval $\delta t$, is $p_i(\delta t)=\sum_{k=1}^{m}\psi_{im+k}/\sum_{j=1}^{K}\psi_j$, usually called the probability measure.  A conventional fractal is adequately described by a single scaling exponent, $D_f$, its fractal dimension, through
$N(\delta t)\sim\delta t^{-D_f}$. 
However, heterogeneous sets require a continuum of scaling indices introduced via the probability $p_i(\delta t)\sim\delta t^{\alpha}$, where $\alpha$ is the strength of the local singularity. Then, $N(\delta t)$ generalizes to $N_\alpha(\delta t)\sim\delta t^{-f(\alpha)}$, where $f(\alpha)$ is the fractal dimension of the subset of intervals characterized by the exponent $\alpha$ \cite{Kad86}.
The non-uniformity of the measure is
captured by the range of multifractality $\theta=\alpha_{max}-\alpha_{min}$,
where $\alpha_{min}$ and $\alpha_{max}$ are the
extreme values of $\alpha$. 
Using another measure $\mu_i(\delta t, q) =
p_i^q/{\sum_j p_j^q}$, where $q$ is a real number, $\alpha$ and $f(\alpha)$ can be directly calculated through
\begin{equation}
\alpha = lim_{\delta t \rightarrow 0} {\frac {\sum_i \mu_i(\delta t, q) ln
p_i(\delta t)}{ln \delta t}},
\end{equation}
and
\begin{equation}
f(\alpha) = lim_{\delta t \rightarrow 0} {\frac {\sum_i \mu_i(\delta t, q)
ln \mu_i(\delta t, q)}{ln \delta t}}.
\end{equation}

\noindent This canonical method was shown to be suitable for the
analysis of short experimental data sets \cite{Chha89}. The
spectrum ($\alpha , f(\alpha)$) is shown in Fig. 5 for $\dot
\epsilon_a = 1.4 \times 10^{-3} s^{-1}$. 
\begin{figure}[h]
\vspace{-2.7cm}
\includegraphics[height=7.5cm,width=5cm]{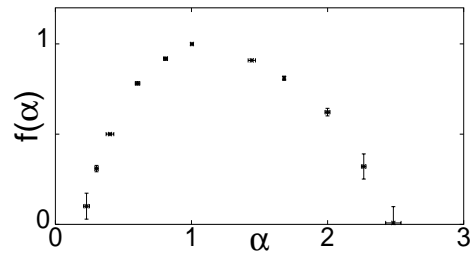}
\vspace{-2cm}
\caption{Multifractal spectrum ($\alpha , f(\alpha)$) for the applied strain rate $\dot{\epsilon}_{a}= 1.4 \times 10^{-3}s^{-1}$, $q \in [-5,+5]$}
\end{figure}
\noindent
The dependence of the multifractal
range $\theta$ on $\dot \epsilon_a$ is shown in Fig. 6. Also
displayed are the different dynamical regimes together with the
band types. $\theta$ is seen to have relatively low values at both
low and high strain rates. The small value of $\theta$ in the
chaotic regime is due to the sharp peak in $D(\Delta \psi)$(Fig. 4a), while that
 in the SOC region is due to the scaling nature of the distribution (Fig. 4c) \cite{Kad86}.
In contrast, a sharp peak  observed at
intermediate $\dot \epsilon_a$ clearly signals a transition
in the nature of dynamics from chaotic B-type to SOC A-type bands.
\begin{figure}[h]
\vspace{-0.3cm}
\includegraphics[height=3.5cm,width=6cm]{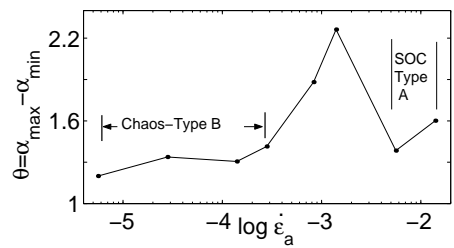}
\vspace{-0.3cm}
\caption{ The multifractal range $\theta$ vs. applied strain rate $\dot{\epsilon}_a$. Regions of chaotic type B and SOC type A bands are marked.}
\end{figure}

Distinct dynamical regimes and spatio-temporal patterns are displayed 
in the above analysis. Chaos is associated 
with type B hopping bands at low applied strain rates and 
SOC dynamics is present at high $\dot\epsilon_a$ in the domain of 
type A propagating bands. The crossover from chaotic to SOC dynamics 
is clearly signaled by a burst in multifractality. Such a diversity in
dynamics poses a challenge for modeling the PLC effect.
What needs to be understood is why the spatial correlations between bands
and the complexity of the dynamics increase with increasing $\dot\epsilon_a$.
The following arguments are quite general and encompass various
mechanisms that have been proposed to explain the spatial coupling
between bands (see \cite{Micha00,Leby95}). A
band of localized deformation induces long range
stresses that favor strain uniformization in its neighborhood. However, 
during jerky flow (i.e., even in the presence of solute atoms)
relaxation of these internal stresses is observed \cite{Chih87}. It occurs by thermally activated rearrangements of dislocation microstructure. 
Thus, both the magnitude and spatial extent of
internal stresses are expected to
decrease with time  once a band  is formed. Let $t_r$ denote this characteristic relaxation time.

At very low strain rates, the reloading time between two successive drops,
$t_l$, is very large and $t_l >> t_r$. The internal stresses are fully 
relaxed and no spatial correlation between bands is observed. This 
corresponds to the domain of type C bands, which are not studied here. 
The absence of spatial correlation 
should make it easier to build a model based on the 
temporal mechanism of the instability, namely the repeated pinning and 
unpinning 
of dislocations in the field of solute atoms. 

At larger strain rates, the reloading time $t_l$ decreases and becomes
commensurate with the relaxation time $t_r$. 
Internal stresses are not
totally relaxed and favor the formation of new bands nearby the previous
ones, hence the hopping character of the associated type B bands.
At high $\dot\epsilon_a$, $t_l << t_r$ very little plastic relaxation occurs
between the stress drops. Further, the reloading rates become increasingly
a significant fraction of the 
unloading rates. Thus, the stresses felt by
dislocations always remain close to the threshold for unpinning from the
solute atoms. New bands can be formed in the field of unrelaxed internal stresses
and recurrent partial events can overlap resulting in a hierarchy of
length scales. This leads to both SOC dynamics and type A propagating bands.

Models incorporating the basic instability mechanism and a spatial coupling \cite{Micha00,Leby95} seem to reproduce the various types of bands and the corresponding stress drop distributions. This suggests that they have the right basis for reproducing the corresponding dynamical regimes, chaos and SOC, as well as the crossover between them.

In summary, a connection is made between the spatial patterns and
the dynamical regimes associated with different strain rates in
the jerky flow of polycrystals. A  qualitative interpretation is
proposed in terms of competing mechanisms that operate at different scales, 
local and global, and involve non-local effects.  
The crossover between the hopping bands associated
with chaos and the propagating bands identified
with the SOC state is well detected by a surge in the multifractal
behavior of the distribution of plastic events. To the best of our
knowledge, this is first example where a crossover from localized
to propagating states has been detected using a multifractal
analysis based on purely experimental signals.

Support by  CNRS and JNCASR
under PICS No. 657 is gratefully acknowledged. G. A. and M. L. are
grateful for the support of Metz University for their stays
 during 1998-2000 and G. A. for the partial support of Department of Science and Technology, Government of India.

\end{document}